\documentclass[useAMS]{mn2e}

\usepackage{graphicx}

\def \dcut{\delta_{\rm cut}}

\def\tesc{t_{\rm esc}}
\def\tcr{t_{cr}}
\def \epm {e_\pm}
\def \etapm {\eta_{\pm,1}}
\def \gpm {\gamma_\pm}
\def \gmin {\gamma_{\rm min}}
\def \gpr {\gamma_{\rm pr}}
\def \gampr {\gamma_{\rm pr}}
\def \epr {\eta_{\rm pr}}
\def \gamr {\gamma_{\scriptscriptstyle R}}
\def \lumr {L_{\scriptscriptstyle R}}
\def \npm {n_\pm}
\def \lbec {L_{\scriptscriptstyle \rm  BEC}}
\def \xibec {\xi_{\scriptscriptstyle \rm  BEC}}

\def \ephi {\epsilon_\phi}

\def \bpcn {B_{\rm pc, 9}}
\def \bpc {B_{\rm pc}}
\def \rpc {r_{\rm pc}}
\def \dom {\Delta\Omega}
\def \dlphi {\Delta l_\phi}
\def \dltheta {\Delta l_\theta}
\def \ldf {L_{\scriptscriptstyle \rm \Delta\Phi}}
\def \lmax {L_{\scriptscriptstyle \rm \Delta\Phi}}
\def \lpart {L_{\rm part}}

\def \etar {\eta_{\sss R}}
\def \lprim {L_{\rm pr}}
\def \lpm {L_{\rm \pm}}

\def \edf {\eta_{\scriptscriptstyle \rm \Delta\Phi}}
\def \emax {\Delta E_{\rm max}}
\def \emx {E_{\rm max}}
\def \sbec {S_{\scriptscriptstyle \rm BEC}}
\def \smean {S_{\rm mean}}
\def \smeanz {S_{\rm mean, 0}}
\def \gmax {\gamma_{\rm max}}
\def \ngj  {n_{GJ}}
\def \son {S_{on}}

\def \nucr {\nu_{\rm crv}}

\def \dbec {\Delta_{\scriptscriptstyle \rm BEC}}

\def \mref#1{(\ref{#1})} 
\newcommand\sss{\scriptscriptstyle}

\newcommand{\rlc}{R_{\rm lc}}

  \newcommand\thpc{\theta_{\rm pc}}

\def \rns{R_{ns}}

\def \numin {\nu_{\rm min}}
\def \numax {\nu_{\rm max}}

\title[Energy budget of J1012$+$5307]
{Energy budget of the bifurcated component in the radio pulsar profile
of PSR J1012$+$5307}

\author[J.~Dyks, and B.~Rudak
]{J. Dyks 
and B.~Rudak\\
Nicolaus Copernicus Astronomical Center, Rabia\'nska 8, 87-100, Toru\'n,
Poland\\
}
\begin{document}

\date{Accepted 2013 July 1. Received 2013 June 11; 
in original form 2013 February 25}


\maketitle

\label{firstpage}

\begin{abstract}
The bifurcated emission component (BEC)
in the radio profile of the millisecond pulsar
J1012$+$5307 can be interpreted as the signature of the curvature radiation
beam polarised orthogonally to the plane of electron trajectory.
Since the beam is intrinsically narrow
($\sim\kern-0.7mm 1^\circ$), the associated emission region must be small
for the observed BEC to avoid smearing out by spatial convolution.
We estimate whether the energy available in the stream
is sufficient to produce such a bright feature
in the averaged profile. 
The energy considerations become complicated
by the angular constraints imposed by the width of the microbeam,
and by the specific spectrum of the BEC which is found to have
the spectral index $\xibec \approx -0.9$ in comparison to the index of 
$\xi\approx-2$ for the total profile.
For typical parameters, the luminosity of the BEC is determined 
to be $4\ 10^{25}$ erg/s, 
whereas the maximum-possible beam-size-limited
power of the stream is $\lmax \approx 2\ 10^{29}$ erg/s. 
This implies the minimum energy-conversion efficiency 
of $\edf \approx 2\ 10^{-4}$. The BEC's luminosity does not exceed 
any absolute limits of energetics, in particular, 
it is smaller than the power of primary electron
and/or secondary plasma stream. However, the implied efficiency
of energy transfer into the radio band 
is extreme if the coherently emitting charge-separated plasma
 density is limited to the Goldreich-Julian value.
This suggests that the bifurcated shape of the BEC has macroscopic
origin, however, several uncertainties (eg.~the dipole inclination
and spectral shape)
make this conclusion not firm.
\end{abstract}

\begin{keywords}
pulsars: general -- pulsars: individual: J1012+5307 --
Radiation mechanisms: non-thermal.
\end{keywords}

\section{Introduction}

Bifurcated emission components (BECs)
have so far been observed
in integrated radio profiles of 
J0437$-$4715 (Navarro et al.~1997)
and J1012$+$5307 (Kramer et al.~1998; Dyks, Rudak \&
Demorest 2010, hereafter DRD10).
It has been proposed in DRD10 that these features are produced
when our line of sight crosses a split-fan beam emitted by
a narrow plasma stream flowing along curved magnetic field lines
(see fig.~1 in Dyks \& Rudak 2012, hereafter DR12).
The double-peaked shape of BECs has been attributed to the
intrinsic bifurcation of the extraordinary mode of the curvature radiation
beam in strongly magnetised plasma.
The peaks in the observed BECs approach each other with increasing frequency
at a rate that is roughly consistent with the curvature radiation origin
(fig.~7 in DR12).
In the case of J0437$-$4715, its BEC is considerably merged.
The observed width of this BEC
is consistent with the angular size of the curvature radiation beam
(sect.~3.1.1 in DR12), and the energy contained in the BEC 
is a small fraction of that observed in the full profile.

In the case of J1012$+$5307, however,
the feature is much more pronounced, wider, and well resolved
(Fig~\ref{smean}).
The BEC's peaks are separated by a deep central minimum which reaches
$\sim32$ per cent the BEC's peak flux at $820$ MHz. 
If the shape of this feature is mostly
determined by the intrinsic shape of the curvature radiation microbeam,
the extent of the associated emission region must be small enough
so that the spatial convolution of the curvature emission beams
does not smear out the BEC. 

At high frequencies ($\nu \ga 3$ GHz, see fig.~5 of Kramer et al.~1999),
the BEC of J1012$+$5307 starts to have comparable flux to the the main pulse 
(MP) in the averaged profile of this pulsar. 
Given the extreme narrowness of the curvature radiation
beam ($\sim 1^\circ$ for typically expected parameters), it is worth
to verify
if the energy supplied by the Goldreich-Julian density in such a narrow
stream is sufficient to produce the observed flux of the BEC.

After introducing some energetics-related definitions in Section
\ref{basics} we estimate the BEC's luminosity (Section \ref{luminosity}).
In Section \ref{power} we estimate
the maximum energy flux that can be confined in the 
plasma stream, the width of which is limited by the resolved form of the
BEC. In Sect.~\ref{gil} we compare our result to another published
estimate, and reiterate our main conclusions in Sect.~\ref{conclusions}.

\section{Basics of energetics}
\label{basics}

The radio luminosity of a pulsar beam
cannot be accurately determined, because we do not know
if our line of sight samples representative parts of the beam. 
Without the a priori knowledge of the emission beam and viewing geometry,
we cannot tell how much the observed flux differs from the flux averaged
over the full solid angle of pulsar emission.
The missing information needs to be provided by some model of the beam
and viewing geometry. The simplest model
assumes a uniform emission beam
of solid angle $\dom(\nu)$, with the uniform emissivity
determined by the observed flux $\son(\nu)$ averaged  
within the `on-pulse' interval of pulse longitude.
This implies the pseudo-luminosity:
\begin{equation}
L=d^2\int\dom(\nu)\ \son(\nu)\ d\nu,
\label{lum0}
\end{equation}
where the integration is
within the frequency band of interest (between $\numin$ and $\numax$).
Hereafter, the term `pseudo', which expresses our assumption 
that the measured flux represents the beam-integrated flux,
will be neglected. 
For many pulsars the observed pulse width
does not change with frequency or it changes slowly enough to consider the
solid angle as $\nu$-independent, and to extract $\dom$ from the integrand
(Gould \& Lyne 1998; Hankins \& Rankin 2010).
In our case $\dom$ and $\son(\nu)$ must be determined for the BEC of
J1012$+$5307. 
The BEC's spectrum will be calculated further below for a $\nu$-independent
pulse longitude interval of $35^\circ$, marked in Fig.~\ref{smean}. 
The solid angle $\dom$ 
will accordingly be considered fixed
($\nu$-independent). Eq.~\mref{lum} then becomes:
\begin{equation}
\lbec=d^2\dom\int \sbec(\nu)\ d\nu,
\label{lum}
\end{equation}
where $\sbec$ is the mean flux of the BEC.

The luminosity of eq.~\mref{lum} cannot exceed the maximum power 
which is theoretically
available for the emitting stream:
\begin{equation}
\ldf=e\Delta\Phi_{pc}\ c\ \ngj(r)\ A(r),\\
\label{lmax}
\end{equation}
where $e\Delta\Phi_{pc}\equiv\emx$ is the energy corresponding 
to the potential drop above the polar cap,
$\ngj$ is the Goldreich-Julian density
of the stream, and $A$ is the crossectional area of the stream, measured
at the same radial distance $r$ as $\ngj(r)$.\\
Note that eqs.~\mref{lum} and \mref{lmax} are not independent: the emitting 
area $A$ in $\lmax$ refers to the same emission region 
as the solid angle $\dom$
in $L$. The choice of the emission region simultaneously determines
both $A$ and $\dom$ in these equations. 

The accelerating potential drop is
approximated by the potential difference between the center and the edge
of the polar cap, as derived for a perfectly conducting neutron star
with vacuum magnetosphere and no dipole inclination:   
$\Delta\Phi_{pc} = 6.6\ 10^{12}\ {\rm V}
B_{pc, 12}R_6^3/P^2$, where $B_{pc, 12}$ is the polar magnetic induction in
TG, $R_6$ is the neutron star radius in units of $10$ km and $P$ is the
pulsar period (Goldreich \& Julian 1969).

To compare the BEC's luminosity with
the power given by eq.~\mref{lmax} one can define the efficiency:
\begin{equation}
\edf \equiv \frac{\lbec}{\ldf}
\label{edf}
\end{equation}
which is expected to be much less than unity.
Since the production of coherent radio emission is not understood 
in detail,
it is also useful to compare the BEC's luminosity to the 
power carried by the outflowing stream of particles
(primary electrons and secondary $\epm$-pairs):
\begin{eqnarray}
\lpart & = & \lprim + \lpm =\\
& = & (\gpr + \npm\gpm)\ m c^3\ \ngj(r)\ A(r),
\label{lpart}
\end{eqnarray}
where $\gpr$ denotes the Lorentz factor acquired by an accelerated 
primary electron,  $\gpm$ is the initial Lorentz factor of $\epm$ pairs, 
and $\npm$ is the number of pairs
produced per one primary electron in the cascade which is responsible
for the BEC.
Eq.~\mref{lpart} defines the power carried by the primary electrons
($\lprim=\gpr mc^3\ngj A$) and the secondary $\epm$ plasma
($\lpm=\npm \gpm mc^3\ngj A$). 
We will also use $L_{\pm, 1}$, which is equal to $\lpm$ taken for $\npm=1$. 
Below we also discuss the radio-emission
power $\lumr$, determined by the minimum Lorentz factor $\gamr$ required
for the curvature radio spectrum to extend at least up to the upper
integration limit $\nu_2$ in eq.~\mref{lum}. 
The stream's radio power can be expressed by:
\begin{equation}
\lumr=\gamr\ mc^3\ \ngj(r)\ A(r).
\label{lumr}
\end{equation}
where a contribution due to $\npm$ is neglected, because
it is electromagnetically difficult 
to separate plasma into charge density
exceeding $\ngj$ (Gil \& Melikidze 2010, hereafter GM10). 
For all the afore-described power-related quantities,
we define their corresponding efficiencies in the way analogical to
eq.~\mref{edf}, eg.: $\epr\equiv
\lbec/L_{\rm pr}$, $\eta_\pm \equiv \lbec/\lpm$, 
$\eta_{\pm,1} \equiv \lbec/L_{\pm,1}$,
$\eta_{\sss R} \equiv
\lbec/\lumr$.

\section{Radio luminosity of the BEC}
\label{luminosity}

The distance $d$ to J1012$+$5307 is estimated to $520$ pc (Nicastro et
al.~1995). 

\subsection{Radio flux of the BEC}
\label{rflux}

\begin{figure}
   \includegraphics[width=0.49\textwidth]{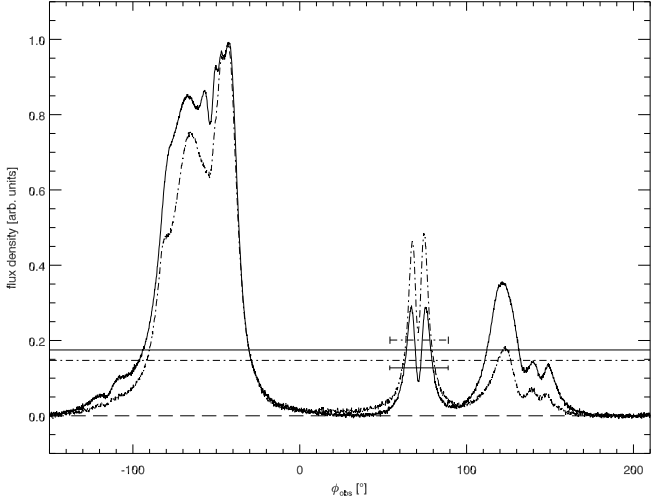}
       \caption{Average pulse profiles and their mean flux 
for J1012$+$5307 at $0.82$ (solid)
and $1.4$ GHz (dot-dashed). The horizontal bars centered at $\phi_{\rm obs}
=71^\circ$
mark the mean flux of the BEC alone, ie.~the flux averaged within 
the $35^\circ$-wide phase interval centered at the BEC. The long horizontal
lines mark the mean flux $\smean$ for the total profiles. The minimum flux near
$\phi=190^\circ$ is assumed to present the zero level (dashed line). 
The profiles are phase-aligned
to match the BECs at both frequencies. This also aligns the trailing side of the
main pulse. Data courtesy: Paul Demorest (DRD10).}  
      \label{smean}
\end{figure}

To estimate $\sbec$, we calculate the mean flux
density within a $35^\circ$-wide interval centered at the BEC, and
compare it with the mean flux density for the total profile
(Fig.~\ref{smean}). This is done by assuming that 
emission from J1012$+$5307 is negligibly low at the pulse longitude
of the minimum flux (around $190^\circ$ in Fig.~\ref{smean}). 
The result, illustrated in Fig.~\ref{smean}, is:
\begin{eqnarray}
\sbec & = & 0.73\ \smean \ \ \rm at\ 0.8\ GHz \label{sona}\\
\sbec & = & 1.37\ \smean \ \ \rm at\ 1.4\ GHz \label{sonb},
\end{eqnarray}
where $\smean = 14$ mJy at $0.8$ GHz, whereas $\smean = 3$ mJy at 1.4 GHz 
(Kramer et al.~1998).
The mean flux density within the BEC is then roughly the same as the
mean flux density of full profile around 1 GHz: $\sbec \approx \smean$.
The case of the BEC of J1012$+$5307 is then considerably different from
the standard case of luminosity estimate for normal pulsars with
narrow beams. In the latter case, $\son$ is an order of magnitude 
larger than $\smean$. This is because to determine the mean flux
$\smean$, the energy contained within the narrow pulse 
is attributed to the full rotation period.
Equation 3.41 in Lorimer \& Kramer (2005, hereafter LK05) is sometimes 
used to quickly estimate radio pulsar luminosities. It is based on the
exemplificative assumption that $S_{\rm peak} = 25\smean$ 
for narrow profiles of typical pulsars (with duty cycle $0.04 = 1/25$).
One should, therefore,
resist from using this equation for the BEC of J1012$+$5307.
In the case of this bifurcated component we have 
$\sbec \sim 0.4 S_{\rm peak}$ (compare the level of bars
at $\phi_{\rm obs}=71^\circ$ in Fig.~\ref{smean} with the peaks of the BEC
at the corresponding $\nu$). The narrow duty cycle expressed by the equation
$S_{\rm peak} = 25\smean$, would then imply $\sbec \sim 10\smean$, 
to be compared with
eqs.~\mref{sona} and \mref{sonb}.
Thus,
if eq.~3.41 from LK05 is directly used to calculate the luminosity of the
BEC, the result becomes overestimated by one order of magnitude.

\begin{figure}
   \includegraphics[width=0.49\textwidth]{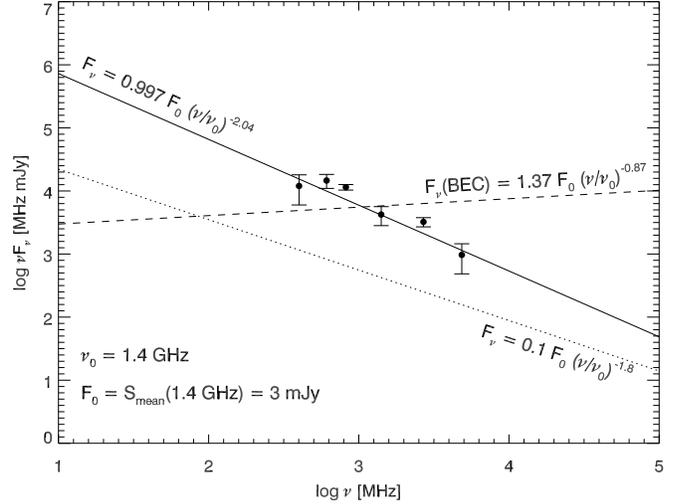}
       \caption{
$\nu F_\nu$ radio spectrum of J1012$+$5307. 
The horizontal axis covers the full range of integration in eq.~\mref{lum}.
Solid line presents the
fitted spectral slope ($\xi = -2.0$). The spectrum
of the BEC (dashed line, normalized according to eq.~\ref{sonb})
is distinct from the total and has $\xi=-0.87$.
The dotted line presents the BEC's spectrum implicitly assumed by GM10.
The dashed spectrum contains
two times more energy than the dotted one. 
Numerical values for all indices are given in the $F_\nu$ convention.  
}
      \label{norma2}
\end{figure}

Kramer et al.~(1999) find that between 0.4 and 5 GHz, the spectra of many MSPs 
can be well approximated by a power-law function of:
\begin{equation}
\smean (\nu) = \smeanz \left(\frac{\nu}{\nu_0}\right)^\xi, 
\label{powerlaw}
\end{equation}
where $\smeanz\equiv\smean (\nu_0)$ is the mean flux density at a frequency $\nu_0$.
To perform the integration in eq.~\mref{lum} the integration limits
are set to $\nu_1 = 10$ MHz and $\nu_2 = 100$ GHz, which is a much larger
interval than has been observationally explored so far for J1012$+$5307.
Data points in Fig.~\ref{norma2} present the $\nu F_\nu$ spectrum of total profile
of J1012$+$5307 based on available literature (Nicastro et al.~1995;
Kramer et al.~1998; Kramer et al.~1999; Stairs et al.~1999). 
Solid line presents the power-law of eq.~\mref{powerlaw}
fitted to the six points in
Fig.~\ref{powerlaw},
with the
index of $\xi = -2.0\pm0.6$ ($\nu F_\nu$ slope of $-1.0$ in the figure).
Using this value of $\xi$, and
the known flux-contributions of the BEC at $0.82$ and $1.4$ GHz
(eqs.~\ref{sona} and \ref{sonb}), we have determined the spectral index
of the BEC $\xibec = -0.87\pm0.58$. The BEC's spectrum, shown
in Fig.~\ref{norma2} with dashed line, is then completely different from
the total one. At $5$ GHz the flux density contained within the BEC
($\sbec$) is $\sim\kern-0.7mm6$ times larger than the mean flux density 
of the total profile. However, the BEC's flux, integrated between 
$\nu_1$ and $\nu_2$, is $14$ times lower than the one calculated for the total
profile with the `total' spectral index of $-2.0$. 
This is caused by
disparate levels of the BEC's and total spectra at low frequencies
(see Fig.~\ref{norma2}).

\subsubsection{Physical implications of the BEC's spectral index}
 
Interestingly, the BEC's spectral index $\xibec =-0.87\pm0.58$ 
is consistent with the value of $-2/3$ expected for distribution of charges
that efficiently loose their energy in the form of the curvature radiation. 
For a narrow (delta-like) source
function, the steady state distribution of particles that undergo the
curvature-radiation cooling has the power-law form $N_e \propto \gamma^{-p}$ 
with $p=4$ (see eq.~3 in Rudak \& Dyks 1999). 
The observed value of $\xibec$ implies
$p=4.6\pm1.7$. 

However, the `curvature spectral index' 
$\xi_{cr} = -2/3$ can extend down to the radio band ($\nu \approx 400$ MHz,
which is the lowest $\nu$ at which the BEC has been detected so far), 
only when the curvature radiation can reduce the electron Lorentz factor down to
$\gamma \sim 40$, as implied by eq.~\mref{nucr} with $\rho=10^6$ cm.
This can occur provided the characteristic time scale of particle escape
from the emission region:
\begin{equation}
\tesc\sim \rho/c \sim 3.3\ 10^{-5}{\rm s}\ \rho_6
\label{tesc}
\end{equation}
is longer than the timescale of the curvature radiation cooling:
\begin{equation}
\tcr\sim \frac{\gamma}{|\dot\gamma_{cr}|} = 
\frac{3}{2}\frac{mc}{e^2\kappa}\frac{\rho^2}{\gamma^3}
=1.2\ 10^{14}{\rm s}\ \kappa^{-1} \frac{\rho_6^2}{\gamma^3},
\label{tcr}
\end{equation}
where $\kappa$ has been introduced to take into account
the increase of energy loss above the vacuum value as a result
of unknown coherency mechanism. The condition $\tesc\ga \tcr$
with $\gamma = 40$
requires $\kappa \ga 5.5\ 10^{13}$, an apparently enormous value. 
Thus, the observed spectrum of the BEC can be understood
as the curvature radiation from an initially-narrow electron energy
distribution, provided that the radiative energy loss rate $\dot\gamma_{\rm
cr}$
is larger by the factor $\kappa$ than the noncoherent value.
However, the uncertainty of $p$ is large, and one cannot exclude the
possibility that other factors are responsible for the observed 
spectral slope.

\subsection{The solid angle}

The value of solid angle $\dom$ depends on the 
beam associated with the observed BEC.
In the `stream-cut' model, the BEC is observed when the line of sight
is traversing through a narrow but elongated, fan-shaped emission beam
(see figs.~1, 2, and 4 in DR12). The transverse width of this beam corresponds
to the angular size of the curvature radiation microbeam.
In what 
follows the word `microbeam' is used to mean the elementary
pattern of radiation characteristic of the coherent emission process
operating in pulsar magnetosphere.
It should be discerned from the `pulsar beam' which is observed at the Earth
and results from spatial convolution of many microbeams.
The BEC of J1012$+$5307, at least around $\nu \sim 1$ GHz, 
appears to be an intermediate case between the
pure microbeam and the convolved case.

\subsubsection{Curvature radiation microbeam}

The beam of curvature radiation emitted in vacuum has a
mostly filled-in, pencil-like shape. 
Deep in the magnetosphere, however, 
the ordinary-mode part of the beam 
can be damped and absorbed by plasma. 
The remaining part, which is the X mode polarised
orthogonal to the plane of electron trajectory, has the two-lobed form
which we associate with the BEC (DRD10). 

The microbeam then consists of two lobes that point at a small angle $\psi$ with
respect to the plane of electron trajectory, with no emission within the
plane itself. The angle between the lobes is:
\begin{equation}
2\psi = \frac{0.8^\circ}{(\rho_7\nu_9)^{1/3}}
\label{microbeam}
\end{equation}
where $\rho = 10^7\ {\rm cm} \times \rho_7$ is the curvature radius of electron
trajectory, and $\nu = 10^9\ {\rm Hz} \times \nu_9$.
Hence for typical parameters 
($\rho_7 \sim 1$, $\nu_9 \sim 1$)
the microbeam size is ten times smaller than the observed separation of
maxima in the BEC around 1 GHz: $\dbec = 7.9^\circ$. We
assume that the large apparent width of the BEC results 
from the very small cut angle $\dcut$ between
the beam and the trajectory of the line of sight.\footnote{Further below we
will discuss wider beams with $\rho_7 \ll 1$ that undergo
less extreme geometrical magnification.}
When we walk across a railway track at a decreasingly small
 angle, 
the distance between two points at which we cross each rail
increases.
Small $\dcut$ increases the apparent width of the BEC in a similar way
(see fig.~2 in DR12, with the angle $\dcut$ marked on the
right-hand side). A BEC produced by the beam of size $2\psi$, effectively 
has the observed width of $\Delta \approx 2\psi/(\sin\zeta\sin\dcut)$,
where $\zeta$ is the viewing angle between the rotation axis and the line of
sight.

There are several important reasons for why
we use eq.~\mref{microbeam} instead of the popular result of 
$\psi \simeq 1/\gamma$, where $\gamma$ is the Lorentz factor of the
radio-emitting electrons: \\
1) Eq.~\mref{microbeam}
is valid for any frequency smaller than, or comparable to, the characteristic
frequency of the curvature radiation spectrum: 
\begin{equation}
\nucr = \frac{3c}{4\pi}\frac{\gamma^3}{\rho} = 7\ {\rm GHz}\ \frac{\gamma^3}
{\rho\ {\rm [cm]}}.
\label{nucr}
\end{equation}
Eq.~\mref{microbeam} does include the result of $\psi \sim 
1/\gamma$ as a special case when $\nu = \nucr$ but it also holds true
for any $\nu \le \nucr$. This can be immediately verified by inserting
$\nucr$ into eq.~\mref{microbeam}, which gives $\psi = 0.78\gamma^{-1}$.\\
2) The approximation given by eq.~\mref{microbeam} is fairly accurate for
frequencies extending all the way up to the peak of the curvature spectrum.
The maximum of this spectrum occurs\footnote{The frequency
$\nucr$ is sometimes defined to be twice larger than in eq.~\mref{nucr}.
In such a case the curvature spectrum peaks at $0.14\nucr$, 
ie.~at a frequency almost one order of magnitude lower than $\nucr$.
The value of $\nucr$ itself is then located at the onset of the exponential 
high-frequency cutoff
of the spectrum, where the flux has already dropped down to 
$\sim\kern-0.7mm30$ per cent 
the peak flux. Note that the \emph{energy} spectrum ($F_\nu$ convention)
is assumed in this discussion of spectral peak location.}
at $\nu_{\rm pk} = 0.28\nucr$. At this spectral peak, eq.~\mref{microbeam} 
overestimates $2\psi$ by only a factor of $1.05$. At $\nu = \nucr$,
the angle is still overestimated only by a modest factor of $1.12$.
\\
3) Unlike eq.~\mref{microbeam}, the formula $\psi \sim 1/\gamma$ is
valid only in two cases: i) for a frequency-integrated BEC;
ii) at the peak of the curvature spectrum: $\nu \simeq \nucr$.
Case ii) does not have to apply for the actual, frequency-resolved
BEC observed at a fixed $\nu$ by a real radio telescope.\\
4) The most important reason: 
within the validity range of eq.~\mref{microbeam}, ie.~for $\nu \le
\nucr$,
\emph{the angle $\psi$ does not depend on $\gamma$}.
Therefore, when the formula $\psi \sim
1/\gamma$ is used instead of eq.~\mref{microbeam}, 
one may misleadingly invoke that for, eg., $\gamma = 10^4$,
the angle $\psi$ at $\nu=1$ GHz is equal to $10^{-4}$ rad $= 0.006^\circ$. 
This is in general wrong, 
because at $\nu = 1$ GHz (fixed by the properties of a radio receiver)
the angle $\psi$ depends on the curvature radius only, and for $\rho=10^7$ cm is
 of the order of  $1^\circ$ \emph{regardless of how high value of
$\gamma$ is assumed.} The curvature radiation has then 
this interesting property
that as long as the curvature spectrum extends up to the
receiver frequency $\nu$, the detected beam has the angular width which 
is fully
determined by the curvature radius $\rho$ only. 
This angular width is with good accuracy 
the same for all values of $\gamma$ that 
ensure
$\nu \le \nucr$.\footnote{It may be worth to note here that in the low
frequency limit $\nu \ll \nucr$, 
the $\nu$-resolved intensity
of the curvature radiation and the shape of the microbeam do not depend on
the Lorentz factor either.}
\\
5) The separation of maxima in the BEC of
J1012$+$5207 evolves with $\nu$ in a way expected in the limit of
$\nu\ll\nucr$ (see fig.~7 in DR12). The use of eq.~\mref{microbeam},
which is also valid in this limit, ensures consistency.\\
6) The formula $\psi \sim 1/\gamma$ is blind to the question of whether
the curvature spectrum for a chosen $\gamma$ extends up to the observed
frequency band. Eg., for $\gamma\sim 10$, which neatly fits the observed
$\dbec$ in the absence of any geometrical magnification, the curvature
spectrum does not reach 1 GHz at all (if $\rho \sim 10^7$ cm). 
Whereas in the case of
eq.~\mref{microbeam} it is immediately visible that an extremely small
$\rho\sim10^4$ cm is required to get $\dbec \simeq 8^\circ$ at 1 GHz.\\

\subsubsection{Value of solid angle}

Since the BEC's beam is emitted by the stream,
the projection of the beam on the sky can be approximated by an elongated
rectangle, described by two dimensions: one in the transverse direction 
orthogonal to plane of the stream (direction of the magnetic azimuth
$\phi$), and the other parallel to the stream (direction of the magnetic
colatitude $\theta$).

The flux contained in the BEC has been estimated in Section \ref{rflux} 
through the integration
over the pulse-longitude interval of $35^\circ$, which is
$4.4$ times larger than the separation of peaks in the BEC at $1$ GHz 
($\dbec \sim 8^\circ$). Since the peak separation itself is 
interpreted as the
angle $2\psi$ given by eq.~\mref{microbeam}, we assume that the transverse 
size of the solid angle is equal to $4.4\times 2\psi = 0.06\ {\rm rad}
/(\rho_7\nu_9)^{1/3}$. The line of sight may cut the beam at a
small angle $\dcut \ll 1$ rad, measured between the elongated projection of
the beam on the sky and the path of sightline passage through the beam.
If $\dcut$ is small whereas the viewing angle $\zeta$ is not, 
the beam needs to extend in the $\theta$ direction 
by an angle comparable to the BEC's phase interval itself ($\sim\kern-1mm 0.5$ rad).
The solid angle associated with the BEC can then be estimated as 
$\dom \simeq 0.03\ {\rm sr} /(\rho_7\nu_9)^{1/3}$. 
This value is similar to the solid angle of a typical
polar beam of normal pulsar (LK05 assume $0.034$ srad).

The BEC's spectrum has been determined above by flux-integration
within the same phase interval at different frequencies.
For consistency, therefore, the solid angle is assumed to be
$\nu$-independent by setting $\nu_9 = 1$, ie.~ $\dom \simeq 0.03\ {\rm srad}
/\rho_7^{1/3}$. 
For the specific spectral index of the BEC ($\xibec \simeq -0.87$, see
Fig.~\ref{norma2})
the resulting luminosity changes only by a factor of $1.2$
if the $\nu$-dependent solid angle is used in eq.~\mref{lum}.
Also note that a choice of wider longitude interval for the BEC does not
change the result much, because the values of $\sbec$ given by
eqs.~\mref{sona} and \mref{sonb} decrease for wider intervals.
This compensates the increase of $\dom$.

\subsection{Luminosity of the BEC}

Taking $d=520$ pc, $\nu_0=1.4$ GHz, $\smeanz = 3$ mJy, 
$\sbec = 1.37 \smeanz$, $\nu_1 = 10$ MHz,
$\nu_2 = 100$ GHz, $\xibec = -0.87$, $\dom = 0.03\
{\rm srad} /\rho_7^{1/3}$ we get:

\begin{equation}
\lbec = 4\ 10^{25}\ \rho_7^{-1/3}\ {\rm erg/s}.
\label{lbec}
\end{equation}

The only previously-known estimate of the luminosity of the BEC
is the one by GM10, 
which has not yet been published
in any astronomical journal, but is being widely broadcasted on most recent
pulsar conferences. The value obtained by GM10 is 15 times larger than
$4\ 10^{25}$. The main reason for this is that GM10 used eq.~3.41 from
LK05, which assumes that because of the {\emph{usually}} narrow duty cycle 
$\delta = 0.04$,
the peak flux $S_{\rm peak}$ is $25$ times larger than 
the mean flux $\smean$.
In the case of the BEC of J1012$+$5307, we have $S_{\rm peak} \simeq 
2.5\smean$ at $\nu \simeq 1$ GHz (see Fig.~\ref{smean}). For the millisecond
pulsars, it is worth to lower down the numerical coefficient in 
eq.~3.41 of LK05 by a factor of $0.04/\delta$, where $\delta$
is the MSP's duty cycle (or to use their eq.~3.40 instead of 3.41).

The other difference is that GM10 used the `global' spectral index of 
the total pulsar population ($\xi=-1.8$) instead of the index of the BEC, 
and assumed that 10\% of the flux
calculated in such a way is contained in the BEC.
Thus, they used the spectrum shown in Fig.~\ref{norma2} with dotted line,
which is different from the BEC's spectrum (dashed line in 
Fig.~\ref{norma2}).
For this reason their estimate of the integral in eq.~\mref{lum} 
is two times \emph{smaller} than ours.
Therefore, GM10 obtain the luminosity which is approximately larger
by a factor $25/2$ than given by eq.~\mref{lbec}.

\section{The maximum power of the stream}
\label{power}

\subsection{Transverse area of the stream}

The transverse crossection of the stream is assumed to extend laterally
through the distance $\dlphi$ (in the direction of magnetic azimuth $\phi$)
and meridionally through $\dltheta$ (in the direction of magnetic colatitude 
$\theta$).
Then the area of the crossection $A = \dlphi\dltheta$.
The size of $\dlphi$ is limited by the spread of magnetic 
field lines within the emitting area $A$, 
which must not be too large in comparison with $2\psi$
(eq.~\ref{microbeam}) to not blur the BEC. 
Let $\theta_B$ denotes the angle between the tangent to a dipolar 
$B$-field line and the magnetic dipole axis. 
For two points separated azimuthally by
$\Delta\phi$, and located at the same $r$ and
$\theta$,
the dipolar $B$-field lines diverge by the angle of $\delta_B$, 
(the angle between
the tangents to the field lines) given by:
\begin{equation}
\sin{\frac{\delta_B}{2}} = \sin{\theta_B}\sin{\frac{\Delta\phi}{2}}.
\label{deltab}
\end{equation}
For emission orthogonal to the dipole axis ($\theta_B = 90^\circ$), 
eq.~\mref{deltab} gives $\delta_B = \Delta\phi$. For two points
on the opposite sides of the polar cap ($\theta_B=1.5\thpc$
and $\Delta\phi=180^\circ$), it gives $\delta_B = 3\thpc\ll\Delta\phi$.
As can be seen, a specific difference $\Delta\phi$ in the magnetic azimuth
results in the $B$-field-line divergence that is smaller 
than $\Delta\phi$ by the factor of
$\sin\theta_B$. This is because for small $\theta_B$,
$B$-field lines become almost parallel to each other (and to the dipole
axis) irrespective of
$\Delta\phi$.  

The divergence $\delta_B$ of the $B$-field lines within the emission region 
is allowed to comprise a fraction
$\ephi$ of the microbeam's width: 
\begin{equation}
\delta_B = \ephi2\psi,
\end{equation}
where $\ephi < 1$ to avoid blurring. Assuming that the allowed angles
$\delta_B$ and $\Delta\phi$ are small, from the last two equations we get:
\begin{equation}
\ephi2\psi = \sin{\theta_B}\Delta\phi.
\end{equation}
This gives the following limitation on the transverse size of the stream:
\begin{equation}
\dlphi = r_\perp\Delta\phi = r_\perp\ephi2\psi/\sin\theta_B,
\label{dlphi}
\end{equation}
where $r_\perp$ is the distance of the stream from the dipole
axis.\footnote{For consistency, $r_\perp$ needs to
refer to the same radial distance $r$ from the star centre,
as the Goldreich-Julian density does in eq.~\mref{lmax}.}
For the rim of the polar cap of J1012$+$5307, we have:
$\theta_B = 17.3^\circ$ which allows $\dlphi$ to be 3.3 times larger than
the value of $r_\perp\ephi2\psi$, expected for orthogonal viewing. 
The BEC of J1012$+$5307 is, however, observed
$~50^\circ$ away from the phase of the interpulse (IP), which may suggest
$\theta_B \sim 50^\circ$, for which $(\sin\theta_B)^{-1} \sim 1.3$.
Since geometric effects can make the observed BEC-IP separation
both smaller and larger than $\theta_B$, below we assume that $(\sin
\theta_B)^{-1}= 1.3$.

\begin{figure}
   \includegraphics[width=0.49\textwidth]{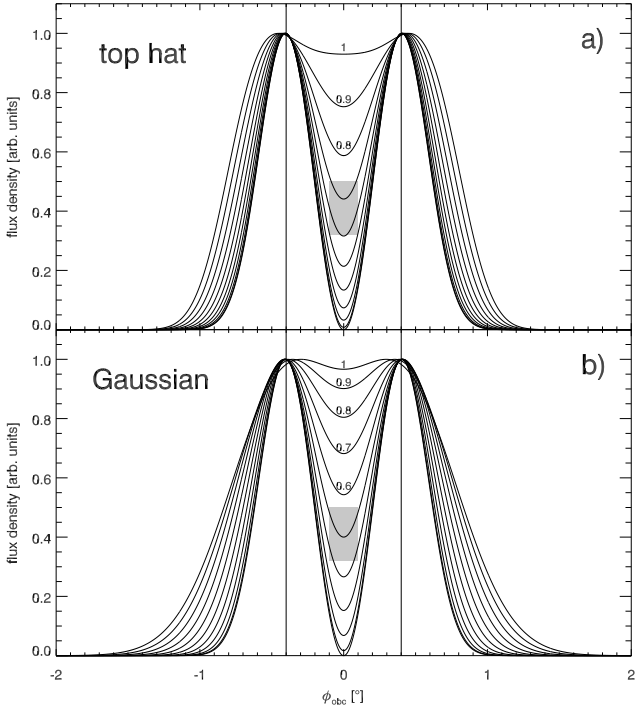}
       \caption{Convolution of the curvature radiation microbeam
with a rectangular distribution of emitting plasma density (a), 
and a Gaussian distribution (b). The lowermost curve in each panel
presents the unconvolved microbeam with
the peak separation of $2\psi$, marked by the vertical lines. 
Different curves correspond to different
widths of a stream $\ephi=1.0, 0.9, 0.8, ..., 0.1, 0.0$ (top to bottom,
a few values of $\ephi$ are marked explicitly for the upper curves), where
$\ephi$ is a fraction of $2\psi$.
The grey rectangle at the center presents the level of flux 
at the center of the BEC of J1012$+$5307,
as measured relative to the BEC's peaks, in the frequency range
 between $0.82$ and $1.4$ GHz (see Fig.~\ref{smean}).
}
      \label{convol}
\end{figure}

The fraction $\ephi$ of the beam size, that can be occupied by the stream
can be estimated by making convolutions of various density profiles
with the shape of the elementary microbeam, given by eq.~(11) in
DR12.

Fig.~\ref{convol} presents such results for
the rectangular (top hat) density distribution (Fig.~\ref{convol}a) and the Gaussian
distribution (Fig.~\ref{convol}b). 
The grey rectangle in the center of both panels presents the observed level
of the central minimum between $0.82$ and $1.4$ GHz: the minimum 
is at 0.32 and 0.5 of the peak flux of the BEC.
In the top hat case this admits the range of $0.6 \le \ephi \le 0.74$.
In the Gaussian case we have assumed that $\ephi=1$ corresponds to the width 
of the Gauss function at the half power level ($1.18\sigma$).
The observed depth of the central minimum then implies $0.44 \le \ephi \le
0.57$. Thus, the maximum allowed width of the stream depends 
on the sharpness of the density distribution. Below we assume
$\ephi=0.5$ at $1$ GHz, 
which approximately corresponds to the Gaussian case in Fig.~\ref{convol}.

If we now consider two points that have the same magnetic azimuth 
$\phi$, but different colatitude $\theta$, we have $\delta_B =
\Delta\theta_B \simeq (3/2)\Delta\theta = (3/2)\dltheta/r$, where 
$r$ is the radial
distance of the emission points from the neutron star
centre.\footnote{Prompted by the referee we explain that
dipolar $B$-field at locations with small magnetic colatitude $\theta$
is inclined at the angle $(3/2)\theta$ with respect to the dipole axis.
Hence the $3/2$ factor.}
This implies $\dltheta$ which is larger than $\dlphi$ by the factor 
of $(2/3)(r/r_\perp)\sin\theta_B$.
However, the spread of $B$-field lines in the colatitude is \emph{not}
limited to a fraction of $2\psi$, because the extent of the emission region
within the plane of the $B$-field lines does not smear out the BEC.
Effects of the colatitude extent are
 degenerate with the effect of motion of electrons along
the $B$-field lines and do not (directly) smear out the BEC.
This can make misleading impression that the colatitudinal extent
is not limited at all by the unsmeared shape of the BEC.
However, because the emission beam is instantaneously narrow,
the parts of the emission region that are far from the line of sight
do not contribute to the detectable flux. This may lead one to think
that $\dltheta$ still needs to be constrained by $\sim\kern-1mm 2\psi$
so that $\Delta\theta_B$ in the stream does not considerably exceed 
the beam size $2\psi$.
This is not the case, because the emission from the poleward
part of the stream (located closer to the dipole axis)
becomes tangent to the line of sight at a slightly larger $r$
(as a result of the curvature of $B$-field lines).
For an arbitrarily large colatitudinal extent $\dltheta$
there exists some radial distance $r$, at
which the poleward extremes of the stream can become visible to the observer. 
Therefore, the extent $\dltheta$ can still 
appear to be unconstrained by the beam size.
However, because of the $r$-dependent effects of
aberration and retardation (Blaskiewicz, Cordes \& Wasserman 1991;
Kumar \& Gangadhara 2012),
the radial extent that is related to $\dltheta$ produces 
a pulse-longitude spread
of $\Delta\phi_{\rm obs} \simeq 2\Delta r/\rlc$
 (Dyks, Rudak \& Harding 2004).
This spread must be a small fraction of the size of the beam: 
\begin{equation}
\Delta\phi_{\rm obs} \simeq 2\Delta r/\rlc \le \epsilon_r 2\psi,
\label{bcw}
\end{equation}
where $\epsilon_r < 1$. For $\epsilon_r = \ephi=0.5$ and $\rlc = 25\
10^6$ cm, this constrains $\Delta r$ to $8.7\ 10^4$ cm.
The extent in colatitude $\dltheta$ is thereby also constrained to a value
that can be determined as follows. Consider the aforesaid two points at the
same $r$ and $\phi$, one of them located at $\theta$, whereas the other
at a slightly smaller colatitude of $\theta(1 - \Delta s)$, 
where $\Delta s \ll 1$. Dipolar field lines are identically
inclined to the dipole axis at all points which 
have the same magnetic colatitude
$\theta$, irrespective of their radial distance $r$.
Therefore, the $B$-field line that crosses the second (poleward) point
becomes tangent to the line of sight at a slightly higher
position $(r+\Delta r, \phi,
\theta)$ determined by the equation of dipolar field lines:
\begin{equation}
\frac{\sin^2\left(\theta(1-\Delta s)\right)}{r} =
\frac{\sin^2\theta}{r+\Delta r},
\end{equation}
which in the small angle approximation gives:
\begin{equation}
\Delta r \simeq 2\Delta s\ r. 
\label{deltar}
\end{equation}
The limit of eq.~\mref{bcw} on $\Delta r$ then translates to:
\begin{equation}
\Delta s \la \frac{\epsilon_r 2\psi}{4}\frac{\rlc}{r}. 
\label{deltas}
\end{equation}
For $\epsilon_r = 0.5$, $2\psi=0.8^\circ= 0.014$ rad, and
$r = 10^6$ cm one obtains $\Delta s  < 0.044$.
Thus, the neccessity to produce the sharply resolved
BEC imposes indirect constraints on the stream's extent in magnetic
colatitude $\theta$:
\begin{equation}
\dltheta = r \Delta\theta = r \Delta s\thinspace \theta \simeq
r_\perp \Delta s \la  \frac{\epsilon_r 2\psi}{4}\frac{\rlc}{r} r_\perp. 
\label{deltat}
\end{equation}
The value of  $\dltheta = (\sin\theta_B/4)(\rlc/r)\dlphi$ may then be
a few times larger than $\dlphi$.
Taking $r_\perp=\rpc$ (rim of the polar cap), one obtains  
$\dltheta  = 0.044\ \rpc$, which is twice as large as
$\dlphi$. 
The apparent bifurcation of the BEC does not therefore put 
equally tight constraints
on the stream size in colatitude, as in the azimuth. 
Actually, it is possible
to consider streams with elliptical crossection, with the longer axis
of the ellipse pointing towards the magnetic pole. Because of the curvature
of magnetic field lines, pair production indeed tends to spread the pairs
in the $\theta$ direction. Let $\Delta\theta_\pm$ denotes the range of
colatitudes over which $e_\pm$-pairs associated to a single primary electron
were produced.
Detailed numerical simulations
of type such as in Daugherty \& Harding (1982)
suggest that $\Delta\theta_\pm$ does not
exceed few hundredths of angular polar cap radius $\thpc$, 
ie.~$\Delta\theta_\pm$ is comparable to $\Delta \theta$ given by eq.~\mref{deltat}.
For the sake of simplicity and minimalism, however, 
we will assume below that
the stream has the same narrow size in both directions: 
$\dltheta =\dlphi$, as given by eq.~\mref{dlphi}.

Using $r_\perp = \rpc = 2\ 10^5$ cm, $\ephi(1\ {\rm GHz})=0.5$, 
and $(\sin\theta_B)^{-1} = 1.3$ in eq.~\mref{dlphi}, we get
$\dlphi=1.8\ 10^3$ cm, and assuming $\dltheta = \dlphi$,
the stream's crossection has the area of
$A = \dlphi\dltheta=3.3\ 10^6$ cm$^2 \rho_7^{-2/3}$. 

\subsection{Kinetic luminosity of the stream}

The electric potential difference between the center and the edge
of the polar cap of J1012$+$5307 is 
$e\Delta \Phi_{\rm pc}=1.45\  10^{14}\ {\rm  eV} = 232$ ergs
which sets up the upper limit to the Lorentz factor:
$\gamma_{\rm max} = 2.8\  10^8$.

Using eq.~\mref{lmax} with the surface value of the Goldreich-Julian
density $\ngj = 4.43\ 10^{18}\ {\rm cm}^{-3}(\dot P/P)^{1/2}$ this
value of $\gamma_{\rm max}$
corresponds to
the following maximum kinetic luminosity of the stream:
\begin{equation}
\lmax = 2\ 10^{29}\ {\rm erg\ s}^{-1}\ \rho_7^{-2/3}.
\label{lmaxres}
\end{equation}
The corresponding value of minimum radio emission efficiency,
calculated using $\lbec$ from eq.~\mref{lbec}, is: 
\begin{equation}
\edf = 2\ 10^{-4} \rho_7^{1/3},
\label{etares}
\end{equation}
which fits the reasonable range expected for radio emission.
Thus, even with the available energy limited by the
narrowness of the beam, the stream has enough energy to power the bright
BEC observed in the average pulse profile of J1012$+$5307.

We have retained dependence on $\rho$, because for $\rho_7=1$ the stream
must be observed at a very small angle ($0.1$ rad) so the intrinsic
beam is enlarged to the apparent BEC. Non-dipolar values of
$\rho_7\ll1$ are, therefore, preferred to make the microbeam wider, and to
place the stream-cut model
in a more comfortable point of the parameter space. This makes the energy
requirements even smaller: for $\rho=10^6$ cm, $\edf = 10^{-4}$,
whereas for $\rho =\rpc = 2\ 10^5$ cm, $\edf = 6\ 10^{-5}$.


Note that except for $\emx$, we have been conservative in our estimates, 
so some
parameters may still be set to make the energy requirements
 even less demanding. For example, the spectrum of the BEC
(Fig.~\ref{norma2})
may be integrated only between $100$ MHz and $10$ GHz, which is already 
a wider interval than has ever been explored for J1012$+$5307.
The stream may be assumed to have an elliptical shape 
with $\dltheta = 0.04\rpc$,
and the `multipolar' $\rho_7=0.1$ may be taken. With all this optimism
applied simultaneously, $\edf = 1.1\ 10^{-5}$.
The energy contained in the BEC is then a negligible fraction of the
maximum energy that can possibly be attributed to the particle stream. 

However, the efficiency is larger when the BEC's luminosity
is compared to the energy of
primary electrons or secondary pairs.  
Let us define
the efficiency: $\eta(\gamma)=\lbec/(\gamma mc^3\ngj A)$.
By setting the upper limit of $\eta(\gamma)=1$ 
one can calculate
the minimum Lorentz factor that the radio-emitting particles
need to have, to supply the energy observed in the BEC:
\begin{equation}
\gmin \simeq 6\ 10^4 (\nu_{100} \rho_7)^{1/3},
\label{gmin}
\end{equation}
where $\nu_{100} = \numax/(100{\rm GHz})$ 
is the upper limit of the frequency-integration in eq.~\mref{lbec}.
The energy transformed into the radio BEC needs to 
outflow at least at a rate exceeding $L_{\rm min} \approx 
\gmin mc^3 \ngj A$.

In the strongly curved $B$-field lines of MSPs, a balance between
the radiative cooling and acceleration will constrain
 the Lorentz factor
of primaries to $9.4\ 10^7 B_{12}^{1/4} P^{-1/8}$ (eg.~Rudak \& Ritter
1994), hence $\gpr = 2.8\ 10^7$. This is clearly larger than $\gmin$,
and implies $\eta_{\rm pr} \simeq 2.2\ 10^{-3}$, ie.~only $0.2$\% of the
primary electron energy is needed to explain the energy of the BEC.
The value obtained in GM10 is $\eta_{\rm pr} \simeq 1$ ($100$\%).

In the case of secondary pairs, their initial
Lorentz factor can be estimated from
the Sturrock's pair condition: $\gpm 
\simeq 9.6\ 10^4 P_{1\rm ms}^{1/2} B_9^{-1}$. For J1012$+$5307,
$\gpm \approx 3.6\ 10^5$ which is six times larger
than $\gmin$. 
Thus, it is enough that only one secondary particle
(out of $n_\pm$ pairs produced per each primary electron) 
transfers $17$\% of its initial energy into the BEC.

However, the secondary electron with so high Lorentz factor
will loose almost all its energy in the form of synchrotron
X-rays, 
not the radio waves. As shown in the appendix,
the remaining energy of parallel motion
is $\gamma_\parallel \approx 25 P_{\rm ms}^{1/2}$, thus
$\gamma_\parallel \approx 57$ for J1012$+$5307. This would have implied 
$\eta_\parallel \equiv \lbec/(\gamma_\parallel mc^3 \ngj A)
\sim 10^3$, however,
such a value of $\gamma_\parallel$ 
is too low for the curvature spectrum to reach the upper limit
$\numax$ of our integration range (if $\rho
\sim 10^7$ cm). This means that either the secondaries are accelerated
or the curvature radius $\rho$ is much smaller than dipolar.
Therefore, to estimate the upper limit for radio emission efficiency,
it is necessary to calculate the minimum Lorentz factor $\gamr$ 
for which the peak 
of CR spectrum reaches the radio band. For 
$\numax = 100$ GHz and $\rho_7=1$, 
eq.~\mref{nucr} gives $\gamr = 522$. 
Hence $\etar \approx 118$.
We emphasize that $\etar$ is independent of curvature radius of electron
trajectory $\rho$, because both $\gmin$ and $\gamr$ 
are proportional to $\rho^{1/3}$. GM10 obtain $\etar \approx 2\ 10^4$ in their
optimistic case, or $\etar \sim 10^6$ for parameters that they call
more realistic.

Thus, although no absolute energy limit is exceeded (there is initially 
$6\npm$ times
more energy in the pair plasma than $L_{\rm min}$),
to explain the observed flux of the BEC,
the available energy would have to be transformed into radio waves
at an extreme rate. 
If the charge-separated (bunched) secondaries loose most of their energy 
in the form of X-rays,
some process is required to draw the energy from another
source, eg.~from the primaries or the charge-unseparated plasma, which
outnumbers the Goldreich-Julian energy flux by the factor of $\npm$.
Alternatively, super-Goldreich-Julian charge densities
would have to emit coherently, ie. $\ngj$ in eq.~\mref{lumr}
would have to be replaced by $n > \ngj$.

\subsubsection{Comparison with the result of GM10}
\label{gil}

Contrary to GM10, we find that the energy content of the BEC does not
break any strict upper limits. Eg.~we find $\edf \simeq 2\ 10^{-4}$, 
$\epr \approx 2\ 10^{-3}$ (in GM10 $\epr \approx 1$), $\etapm \approx
0.17$. For the radio emission efficiency we get 
$\etar \approx 10^2$,
ie.~we confirm the need for extremely efficient energy
transport into the radio band.
However, GM10 using a similar method estimate $\eta_R \approx 10^4 -10^6$, 
which is in a notable disagreement with our result.
There are several reasons for this difference:

1) GM10 have overestimated $\lbec$ by a factor of 15, because 
the duty cycle of J1012$+$5307 is much larger than $0.04$, which is the
typical duty cycle of normal pulsars assumed in LK05.

2) GM10 assume $\gampr=5\ 10^6$, ie.~for unspecified reason
they assume that only $1.8\%$ of available potential drop can be used up
for powering the stream.
We use the maximum Lorentz factor that the primary electrons 
can reach in the
radiation-reaction-limited acceleration.
We emphasize, however, that
the energy available for radio-emitting $e^\pm$ pairs
may actually be larger than $\gampr$, because when primary electrons 
are moving up with a fixed
Lorentz factor (balanced by the energy losses to the curvature radiation), 
the energy is anyway being produced in the form of curvature photons 
that can produce the radio-emitting electron-positron plasma.
It is then possible to produce the energy in form of the electron-positron 
plasma without any change of electron energy 
(or even while the electron energy is increasing).
For this reason the energy
available for the stream may have more to do with the maximum potential drop
rather than with the maximum achievable Lorentz factor.

3) GM10 neglect the factor $(\sin\theta_B)^{-1}$ in eq.~\mref{dlphi},
which at the polar cap's rim of J1012 can 
increase the maximum allowed width of 
the stream $3.3$ times. We assume $\theta_B= 50^\circ$ and 
$(\sin\theta_B)^{-1}=1.3$.

4) Furthermore, GM10 suggest that $\lmax$ should be decreased by a factor of
$7$ because the double-lobed, orthogonally-polarised part of the curvature
microbeam comprises only $1/7$ part of the total energy contained
in the vacuum curvature beam. 
However, the form in which
the energy of the parallel mode leaves the stream is not obvious.

5) GM10 assume
the Lorentz factor of radio emitting plasma $\gamma_\pm=400$, 
the stream size
of $\ephi\times(1/\gamma_\pm)$, and $\ephi=0.1$ 
to obtain $L_{\sss R}$ which is several
orders of magnitude lower than $\lbec $. 
However, their set of parameters
($\gamma_\pm=400$, $\ephi=0.1$) is self-inconsistent because
the BEC has the well-resolved double form at $\nu\simeq 1$ GHz, whereas
for typical $\rho_7=1$, the value of $\gamma_\pm=400$ 
corresponds to $\nucr=45$ GHz, which is in the range where 
the BEC is unresolved and $\ephi$
can considerably exceed $1$.
Around $1$ GHz the flux observed at the minimum between the peaks 
increases quickly up and it is reasonable to expect that the BEC is fully
merged at $\nu \gg 1$ GHz. Thus, the condition $\ephi=0.1$
may apply only for $\nu \la 1$ GHz, whereas at high $\nu$ the stream
may well be wider than the microbeam ($\ephi > 1$).

The BEC is well-resolved around $1$ GHz, and the formula for
the beam size used by GM10 ($\psi \sim1/\gamma_\pm$) is only valid at 
$\nu\simeq \nucr$. Therefore, both $\ephi$ and the beam size 
(hence, the value of $\gamma_\pm$)
must refer to $\nu\simeq 1$ GHz. However, 
the large value of $\gamma_\pm=400$ can
only be made consistent with $\nucr = 1$ GHz if $\rho = 448\ 10^6$ cm
(as implied by eq.~\ref{nucr} with $\gamma_\pm=400$ and
$\nucr=1$ GHz). The value of $\rho$ implicitly present in 
their beam-size calculation is $\rho = 18\rlc$, where $\rlc=25\ 10^6$
cm is the light cylinder radius. 
Thus, to justify their pessimistic values of $L_{\sss R}$,
GM10 assume parameters that imply the curvature radius 
several times larger than the light cylinder radius.
This should not be practiced to 
find the \emph{maximum available} kinetic luminosity.

6) GM10 increase $\gamma_\pm$
to decrease the stream crossection $A\propto (1/\gamma_\pm)^2$, 
while keeping the BEC's luminosity
fixed, which again implies $\eta_R \gg 1$. One should remember, however,
that the beam size also determines the size of the solid angle $\dom$
in eq.~\mref{lum} for the BEC's luminosity. 
For the narrowing split-fan beam, our line of sight
must cut it at a smaller angle $\dcut$ 
so that the $8^\circ$-wide BEC is observed. The value of the solid angle
$\dom$, which is proportional to the beam's width $2\psi$ 
(or, in the case of GM10, to $1/\gamma_\pm$), should therefore 
be decreased accordingly. 
For $\dltheta\propto\dlphi\propto \psi$ 
the ratio $\lbec/L_{\sss R}$ is then proportional to
$\psi^{-1}$ 
instead of being proportional to $\psi^{-2}$.
It is then neccessary to treat
eqs.~\mref{lum} and \mref{lmax} as related to the same beam opening
angle to avoid the overestimate of $\eta_R$.

\section{Conclusions}
\label{conclusions}

We have calculated the radio luminosity of a single component
selected from an average radio pulsar profile: 
the bifurcated component of J1012$+$5307.
This luminosity, equal to $4\ 10^{25}$ erg/s has been attributed to a
narrow stream of radio-emitting plasma. The width of this stream is limited
by the opening angle of the curvature radiation microbeam at $1$ GHz,
as determined by the well-resolved double-peak form of the BEC at this
frequency. It has been shown that the efficiency of converting the
stream's energy into the BEC's radio luminosity is of the order of
$\edf\approx 2.2\ 10^{-4}\rho_7^{1/3}$, 
$\epr \approx 2\ 10^{-3}$, $\etapm \approx 0.17$, when the cross-cap
potential drop, maximum energy of primary electrons, or initial energy
of pairs is taken as a reference, respectively.
Thus, no absolute energy limits are violated,
and there is no energy deficit that can definitely be considered `fatal'
(as phrased by GM10)
for the microbeam model of the BEC.

However, this result implies that large fraction of initial energy 
of a single secondary 
electron (per one primary) needs to be transferred to the radio BEC.
This is unlikely, because such pairs loose most of their energy in the form of 
synchrotron X-ray photons. To power
the radio emission, therefore, the energy would have to be drawn from the
primary electrons or from the remaining 
charge-unseparated plasma (there is an extra energy of 
$\npm-1$ unbunched secondary particles per each primary).
Instead of assuming that such processes occur, it may be more natural
to conclude that the BEC of 1012$+$5307 has macroscopic origin.

\begin{figure}
   \includegraphics[width=0.49\textwidth]{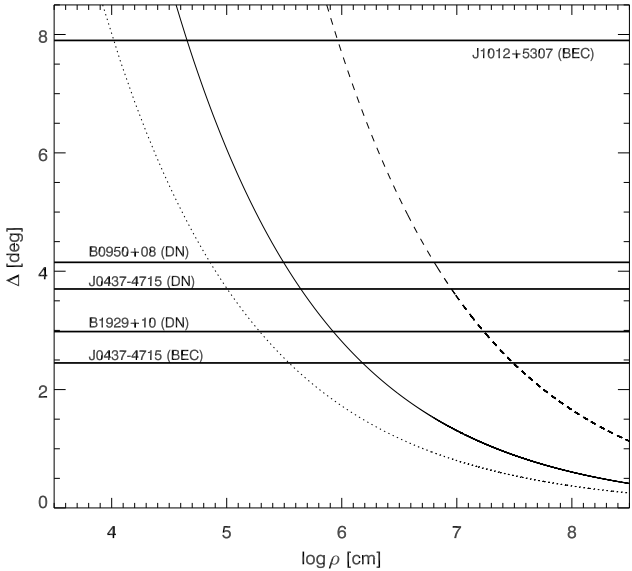}
       \caption{Peak separation $\Delta$ of pulsar double features 
as measured at 1 GHz (horizontal lines). The solid curve presents
the relation $\Delta(\rho)=2\psi/(\sin\zeta\sin\dcut)$ 
for the viewing angle 
$\zeta=60^\circ$, stream-cut angle $\dcut=45^\circ$,
and $2\psi$ given by eq.~\mref{microbeam} (curvature radiation).
The other curves have
$(\zeta, \dcut) = (90^\circ, 90^\circ)$ (dotted), 
and $(60^\circ, 15^\circ)$ (dashed).
Most features are consistent with $\rho \la 10^6$ cm for statistically
average geometry (solid curve).
Note the isolated location of J1012$+$5307. 
}  
      \label{size}
\end{figure}

There are indeed some aspects that make the BEC of J1012 different from
the rest of double features: 1) it is much wider, see Fig.~\ref{size},
and 2) the outer wings of the BEC are much less steep than in the double
notches of B1929$+$10 (see section 3.3 in DR12).

Nevertheless, the locally bidirectional emission (whether of either 
micro- or macroscopic origin) remains a valid and successful model 
for the BEC of
J1012$+$5307. Note that it was not the BEC of J1012, but
the absolute depth of double notches, which has decisively supported the model 
of bidirectional curvature radiation
in section 4 of DRD10 (for physical details on the beam see 
Gil, Lyubarsky \& Melikidze 2004). 
The curvature microbeam model continues to remain a valid and successful 
explanation for 
all the other pulsar double features (DR12). 

We have assumed throughout this paper that
J1012$+$5307 is a highly-inclined rotator, with a large dipole inclination 
$\alpha\sim90^\circ$, and a large viewing angle $\zeta$, measured between
the star's rotation axis and the line of sight.
 The orthogonal geometry is supported by the presence of the 
interpulse separated by half of the rotation period from the MP,
as well as by the width of the MP itself (about $40^\circ$),
which is very close to the opening angle of the surface polar beam
($35^\circ$).
For small $\alpha$ and $\zeta$, which we consider unlikely,
both the luminosity of the BEC,
and the radio efficiency $\eta_R$ become smaller than quoted above.
This is because the observed width of the BEC 
(a few times larger than $\dbec$)
 corresponds to the intrinsic
solid angle $\dom$ that becomes smaller by a factor of $\sin^2{\zeta}$.

Our results can additionally be affected by
the uncertainty in the spectral index, bulk shape of spectral energy
distribution (spectral breaks or cut-offs in the yet-unexplored frequency
range), distance and scintillation-affected mean flux.
However, since most of these factors can bias the result in both directions,
it is unlikely that they can considerably decrease the energy requirements.

The value of the polar cap radius $\rpc$ that enters eqs.~\mref{dlphi}
and \mref{deltat} is known at best with the accuracy $16$\%, 
which is the difference between the vacuum and force free case
(fig.~4 in Bai \& Spitkovsky 2010). This implies a $36\%$ uncertainty in the
stream area $A$. If a more narrow spectral band is assumed
($0.1$ -- $10$ GHz instead of $0.01$ -- $100$ GHz) the BEC's luminosty estimate
decreases by a factor of $2.1$. However, widening the band up to the range
$(1\ {\rm MHz}\  - 1000\ {\rm  GHz})$ increases $\lbec$ by only a factor of
$1.6$.
If the spectral index of the BEC is increased or decreased by $0.5$ 
the BEC's luminosity increases by a factor of $\sim 2$.
This is because in both these cases the spectrum becomes steep
in comparison to the present slope of $+0.12$ in the $\nu F \nu$ convention.

The possibility of spectral breaks beyond the GHz band can reduce the
energy requirements. Most millisecond pulsars do not exhibit
any spectral breaks above 100 MHz (Kuzmin \& Losovsky 2001, hereafter KL01;
Kramer et al.~1999)
For J1012$+$5307 however, the Pushchino measurements
at 102 MHz (KL01; Malofeev et al.~2000)
suggest a break at $\sim600$ MHz. This break is not included in our analysis
(and not shown in Fig.~\ref{norma2}), because the BEC is not discernible 
in the low-frequency profiles (see fig.2 in Kondratiev et al.~2012).
It is therefore not possible to determine the fractional energy
content of the BEC below the spectral break. However, for the BEC
spectrum shown in Fig.~\mref{norma2} with the dashed line, the mean flux
of the BEC becomes comparable to the mean flux of total profile
near $100$ MHz (30 mJy, KL01). The lack of strong BEC 
in the $100$ MHz profiles implies that the actual spectrum 
of the BEC at low frequencies is steeper than in the GHz range.

The Goldreich Julian density in the emission region, 
which is proportional to $\vec \Omega \cdot \vec B$ 
can be much smaller than we assume, if the local $\vec B$ 
is orthogonal to $\vec \Omega$. This can happen because
we assume rather large viewing angle. However, although we fix
$\zeta$ to $90^\circ$ for practical reasons, any values in the broad
range of $\zeta = 90^\circ \pm 45^\circ$ are possible.
Moreover, local non-dipolar enhancements of $B$ are capable of increasing 
the density to a level considerably higher than the dipolar one.
If the dipole inclination is not orthogonal, the magnetic field
derived from the dipolar radiation energy loss should furthermore
be increased by the factor $(\sin\alpha)^{-1}$. Other breaking mechanisms
introduce additional uncertainty. For example, the magnetospheric currents
in PSR B1931$+$24 change $\dot P$ by a factor of $1.5$ (Kramer et al.~2006), 
which implies a $22$\% error in $B$.

Contrary to GM10 we find that the BEC's flux comprises a tiny part
of the maximum limit for the stream energy 
($\epr \approx 2\ 10^{-3}$, as compared to $\epr\approx 1$ in GM10).
The radio emission effciency is indeed extreme (we find $\etar \approx 10^2$),
yet it is 2--4 orders of magnitude smaller than in GM10.  
The determination of luminosity and efficiency for isolated components 
in pulsar profiles is more complicated than standard energy considerations.
Special care is required because:
1) Such an isolated component may have considerably 
different spectrum than 
the total
pulsar spectrum and it may have the flux density which can be at any ratio
with the mean flux density of the total profile.
2) Fraction of the polar cap outflow that is responsible for the observed
component needs to be carefully estimated, with the constraint of not
blurring the component which is observed sharp and resolved
at a frequency $\nu$. In the case of the split-fan beam, this `sharp view'
condition is different in the transverse ($\phi$-) direction
than in the meridional direction of $\theta$.
3) For $\nu \le \nucr$, the size of the curvature 
microbeam \emph{at a fixed frequency 
$\nu$} is independent of the Lorentz factor $\gamma$ 
of the radio emitting plasma. 
The popular formula $\psi \sim \gamma^{-1}$ can only be used
when $\nucr(\gamma, \rho) = \nu$, where $\nu$ is the center frequency
of the observed bandwith. If the ratio $\nu/\nucr$ is not known,
and the spectrum extends up to the observation band,
it may be more safe to use eq.~\mref{microbeam}.
4) Both the radio luminosity of the BEC, through $\dom$, 
as well as the maximum power of the stream, through $A$,
depend on the width of the microbeam. Any decrease of available power
imposed by the decreasing width of the beam is alleviated by 
the simultaneous decrease of BEC's luminosity.

\section*{Appendix}
 
The energy of parallel motion of $\epm$ pairs can be estimated
in the following way. Let us consider a secondary electron with initial
Lorentz factor $\gamma_\pm$.
Let $v_\parallel$ be the component of this electron's velocity
parallel to the magnetc field, and
$\gamma_\parallel=(1-(v_\parallel/c)^2)^{-1/2}$ is the corresponding
Lorentz factor. Now consider a primed Lorentz frame which moves along 
$\vec B$ with the velocity $v_\parallel$. In this frame our electron has
purely transverse velocity $v^\prime_\perp$ and a Lorentz factor
$\gamma^\prime_\perp=(1-(v^\prime_\perp/c)^2)^{-1/2}$. The Lorentz
transformation of velocities implies:
\begin{equation}
\gamma_\parallel = \frac{\gamma_\pm}{\gamma^\prime_\perp}.
\label{iloczyn}
\end{equation}
In the case of millisecond pulsars,
the Sturrock's condition for pair creation
is:
\begin{equation}
\chi \equiv\frac{1}{2}\frac{\epsilon}{mc^2}\frac{B}{B_Q}\sin\psi \approx
\frac{1}{11.5},
\label{sturrock}
\end{equation}
(Sturrock 1971), where $\epsilon$ is the energy of the pair-producing photon, 
$mc^2$ is the
electron rest energy, $B_Q\approx44$ TG is the critical magnetic field value,
and $\psi$ is the angle at which the photon crosses the magnetic field.
For classical pulsars the number on the right hand side is closer to
$1/15$.
In the case of $\chi \ll 1$, each component of the created pair takes up half
energy of the parent photon: 
\begin{equation}
\gamma_\pm mc^2 \approx \frac{\epsilon}{2}
\label{halfen}
\end{equation}
(Daugherty \& Harding 1983) and follows the photon's propagation direction.
In the relativistic limit of $\gamma_\pm \gg 1$, it holds that 
$\cos\psi = v_\parallel/v_\pm \approx v_\parallel/c$, hence:
$1/\gamma_\parallel \approx \sin\psi$. Eqs.~\mref{sturrock} and
\mref{halfen} then give:
\begin{equation}
\gamma^\prime_\perp = 3.8\ 10^3 \left(\frac{r}{\rns}\right)^3
B_{pc, 9}^{-1},
\label{gamperp}
\end{equation}
where the local $B$ field is $B = B_{pc}(r/\rns)^{-3}$.

The Lorentz factor $\gamma_\pm$ can also be estimated from 
\mref{halfen} and \mref{sturrock}, by noting that a photon emitted
in dipolar field at $(r, \theta)$ encounters the largest value of 
$B\sin\psi = 0.085\theta B(r)$ (Rudak \& Ritter 1994).
This is approximately the place where the one-photon absoption coefficient
is maximum, and the pair production is most likely and efficient.
By inserting the last formula into \mref{sturrock} one can derive 
so called `escape
energy', which is the minimum photon energy required to produce
pairs in pulsar magnetosphere: 
\begin{equation}
\epsilon_{\rm esc} \approx 1.0\ 10^5 \ {\rm MeV}\ 
R_{ns, 6}^{-1/2} (r/\rns)^{5/2} B_{pc, 9}^{-1} P_3^{1/2},
\label{eesc}
\end{equation}
where $R_{ns, 6} = \rns/(10^6\ {\rm cm})$, $P_3 = P/(10^{-3}\ {\rm s})$,
and $\theta\approx (r/\rlc)^{1/2}$ was assumed to correspond 
to the polar cap rim.
By inserting \mref{eesc} into \mref{halfen} we get:
\begin{equation}
\gamma_\pm = 9.4\ 10^4 (r/\rns)^{5/2} P_3^{1/2} B_{pc, 9}^{-1}.
\label{gampm}
\end{equation} 
From \mref{iloczyn}, \mref{gamperp}, and \mref{gampm} we obtain
\begin{equation}
\gamma_\parallel = 25\ (r/\rns)^{-1/2} P_3^{1/2},
\label{gampar}
\end{equation} 
which, in the limit of near-surface emission and pair production
($r\sim\rns$) is used in the main text. 
The derived estimates well reproduce the
results of exact numerical simulations (see the 
distributions of $\gamma_\parallel$
and $\gamma^\prime_\perp$ for a normal and millisecond pulsar 
in fig.~1 of Rudak \& Dyks 1999). They are also useful 
in semi-analytical modelling of pair cascades (Zhang \& Harding 2000).

\section*{acknowledgements}
We thank Paul Demorest for providing
us with GBT data on J1012$+$5307.
This work was supported by 
the National Science Centre grant DEC-2011/02/A/ST9/00256
and the grant N203 387737
of the Polish Ministry of Science and Higher Education.

\section*{REFERENCES}

Bai X.-N., Spitkovsky A., 2010, ApJ, 715, 1282\\
Blaskiewicz M., Cordes J.M., Wasserman I., 1991, ApJ, 370, 
643\\
Daugherty J.K., Harding A.K., 1983, ApJ 273, 761\\
Rudak, B. \& Dyks, J., 1999, MNRAS, 303, 477\\
Dyks J., \& Rudak B., 2012, MNRAS, 420, 3403 (DR12)\\
Dyks J., Rudak B., \&  Demorest P., 2010, MNRAS, 401, 1781
(DRD10)\\
Dyks J., Rudak B., \& Harding A. K., 2004, ApJ 607, 939\\
Gil J.A., \& Melikidze G.I., 2010, astro-ph/1005.0678
(GM10)\\
Gil J., Lyubarsky Y., \& Melikidze G.I., 2004, ApJ, 600,
872\\
Goldreich P., \& Julian W.H., 1969, ApJ, 157, 869\\
Gould, D.M., \& Lyne, A.G. 1998, MNRAS, 301, 235\\
Hankins, T.H., \& Rankin, J.M., 2010, ApJ, 139, 168\\
Kondratiev V., Stappers B., \& the LOFAR Pulsar Working
Group, 2012, Proc.~of the IAU Symposium 291, ed.~J. van Leeuwen, Cambridge
University Press, 2012, 47\\
Kramer, M., Lyne, A.G., O'Brien, J.T., et al.~2006, Science,
  312, 549\\
Kramer M., Xilouris K. M., Lorimer D., Doroshenko O.,
Jessner A., Wielebinski R., Wolszczan A., Camilo, F. 1998, ApJ, 501, 270\\
Kramer M., Lange C., Lorimer D., Backer D.C., 
Xilouris K. M., Jessner A., \& Wielebinski R. 1999, ApJ, 526, 957\\
Kumar D., \& Gangadhara R.T. 2012, ApJ, 746, 157\\ 
Kuzmin A.D., \& Losovsky B.Ya. 2001, A\&A, 368, 230 (KL01)\\
Lorimer D.R., \& Kramer, M., 2005, {\it "Handbook of pulsar
astronomy"}, Cambridge University Press, Cambridge (LK05)\\
Malofeev V.M., Malov O.I., \& Shchegoleva N.V., 2000,
Astron.~Rep., 44, 436\\
Navarro J., Manchester R. N., Sandhu J. S., 
Kulkarni S.R., Bailes M., 1997, ApJ, 486, 1019\\
Nicastro L., Lyne A.G., Lorimer D.R., et al., 1995, MNRAS,
 273, L68\\
Rudak, B., \& Dyks, J. 1999, MNRAS, 303, 477\\
Rudak, B., \& Ritter, H., 1994, MNRAS, 267, 513\\
Stairs I.H., Thorsett S.E., Camilo F., 1999, ApJSS, 123, 627\\
Sturrock, P.A. 1971, ApJ, 164, 529\\
Zhang, B., \& Harding, A.K. 2000, ApJ, 532, 1150\\


\end{document}